\documentclass[aps,showpacs,preprintnumbers,amsmath, amssymb]{revtex4}

\usepackage{float}
\usepackage{graphics,epsfig}
\usepackage{graphicx}
\usepackage{dcolumn}
\usepackage{bm}

\begin{document}
\baselineskip=0.8 cm
\title{{\bf A general holographic insulator/superconductor model with dark matter sector away from the probe limit}}
\author{Yan Peng$^{1,2}$\footnote{yanpengphy@163.com}}
\affiliation{\\$^{1}$ School of Mathematical Sciences, Qufu Normal University, Qufu, Shandong 273165, China }
\affiliation{\\$^{2}$ School of Mathematics and Computer Science, Shaanxi Sci-Tech University, Hanzhong, Shaanxi 723000, China }
\author{Qiyuan Pan$^{3}$\footnote{panqiyuan@126.com}}
\affiliation{\\$^{3}$ Department of Physics, Key Laboratory of Low
Dimensional Quantum Structures and Quantum Control of Ministry of
Education, Hunan Normal University, Changsha, Hunan 410081, China}
\author{Yunqi Liu$^{4}$\footnote{liuyunqi@hust.edu.cn}}
\affiliation{\\$^{4}$ School of Physics, Huazhong University of
Science and Technology, Wuhan, Hubei 430074, China  }

\vspace*{0.2cm}
\begin{abstract}
\baselineskip=0.6 cm
\begin{center}
{\bf Abstract}
\end{center}

We investigate holographic phase transitions with dark
matter sector in the AdS soliton background away from the probe limit.
In cases of weak backreaction, we find that the larger coupling parameter $\alpha$ makes
the gap of condensation shallower and the critical chemical potential keeps as a constant.
In contrast, for very heavy backreaction, the dark matter sector could affect the critical
chemical potential and the order of phase transitions.
We also find the jump of the holographic topological entanglement entropy
corresponds to a first order transition between superconducting states in this model with dark matter sector.
More importantly, for certain sets of parameters, we observe novel phenomenon of retrograde condensation.
In a word, the dark matter sector provides richer physics in the phase structure
and the holographic superconductor properties are helpful in understanding dark matter.

\end{abstract}

\maketitle
\newpage
\vspace*{0.2cm}

\section{Introduction}

The anti-de Sitter/conformal field theories (AdS/CFT) correspondence
relates the strongly correlated conformal field theory on the
boundary with a weakly interacting gravity system in the bulk
\cite{Maldacena,S.S.Gubser-1,E.Witten}. According to this novel idea,
the gauge/gravity duality has been successfully employed to gain a
better understanding of the low energy physics in condensed matter
systems from a higher dimensional gravitational dual. The simplest
holographic metal/superconductor model dual to gravity theories was
constructed by applying a scalar field and a Maxwell field coupled
in the AdS black hole background \cite{S.A. Hartnoll,C.P.
Herzog,G.T. Horowitz-1}. And then the holographic
insulator/superconductor transiton was also established in the AdS
soliton spacetime \cite{TN,GTH}. At present, a lot of more
complete holographic superconductor models have been widely studied
to model conductivity and other condensed matter physics properties
in various gravity theories, such as the Einstein-Gauss-Bonnet
gravity, Horava-Lifshitz gravity, non-linear electrodynamics gravity
and so on, see Refs. \cite{R}-\cite{MB}.

According to contemporary astronomical observations, almost 24
percent of the total energy density in our universe is in the form
of dark matter, whose configuration is still unclear. In order to
model the dark matter, a new gravity was established by introducing
an additional U(1) gauge field coupled to the normal Maxwell field
\cite{HD,TA}. This dark matter gravity is strongly supported by
astrophysical observation of 511 keV gamma rays \cite{SP1} and the
electron positron excess in galaxy \cite{SP2,SP3}. Moreover, this
gravity reveals new physics allowing the $3.6\sigma$ discrepancy
between measured value of the muon anomalous magnetic moment and its
prediction in the Standard Model \cite{SP4}. In order to understand
dark matter from properties of holographic systems, new holographic
metal/superconductor transition models were constructed in the
background of this dark matter gravity, for references see
\cite{LN-1,LN-2}. The interesting topics of effects of dark matter
on holographic vortices and holographic fluid viscosity were also
carried out in \cite{MRKW2,MRKW3}. Since the holographic
superconductor provides a lot of qualitative characteristic
properties shared by real superconductor, it is believed that
holographic superconductor theory may be applied to superconductor
in the laboratory. As a further step along this line, some possible
methods to directly detect the dark matter in the lab were proposed
according to holographic theory \cite{MRKW4}. Another interest to
study this model is related to the question of possible matter
configurations in AdS spacetime \cite{TR,BM}. On the sides of CFT on
the boundary, it was shown that the dark matter sector really
provides richer physics in the critical temperature and also the
stability of metal/superconductor transitions on the boundary. Most
surprisingly, for a certain set of parameters, there is a novel
phenomenon of retrograde condensation in holographic dark matter
sector found in Refs. \cite{LN-1,LN-2}, which was already observed
in other metal/superconductor systems \cite{FD,RL}. We further
obtained the general conditions for retrograde condensations and
proved retrograde condensations to be unstable in the s-wave
metal/superconductor transitions \cite{Yan Peng-0}. In this work, we
mainly focus on the effects of the dark matter in holographic
insulator/superconductor model, which is interesting because the
method provides the insight into the strong coupling system and
possibly it enables findings of some `experimental' facts connected
with dark sector.

Most of holographic models were constructed in the background
of AdS black hole or described metal/superconductor transitions. So the holographic
insulator/superconductor transition in the AdS soliton spacetime needs much more research.
Along this line, the dark matter sector was
also considered in the AdS soliton spacetime while neglecting the
matter fields' backreaction on the metric \cite{LN-3,LN-4,MRKW1}.
It was shown that the critical chemical potential is independent of the
dark matter sector parameter without backreaction.
In this work, we will show that the dark matter sector could affect
the critical chemical potential and also the order of transitions when including backreaction.
In contrast to retrograde condensation in
holographic metal/superconductor models \cite{LN-1,LN-2,FD,RL}, we will also manage to obtain the novel
retrograde condensation in our s-wave holographic insulator/superconductor model.

On the other hand, the entanglement entropy is usually applied to keep
track of the degrees of freedom for strongly coupled system while other traditional methods might not be available.
According to the AdS/CFT correspondence, it was proposed that the holographic entanglement entropy of a
strongly interacting system on the boundary can be calculated from
a weakly coupled gravity dual in the bulk \cite{S-1,S-2}.
With this elegant approach, the holographic entanglement entropy has recently been
used to study properties of phase transitions in various holographic models
and it provides us richer insights into the phase transitions \cite{NishiokaJHEP}-\cite{Cai-5}.
For example, it was argued that the entanglement entropy is a good probe to the
critical phase transition point and also the order of phase transitions.
As a further step, we would like to
examine whether the holographic entanglement entropy is still useful
in disclosing properties of transitions in this general holographic
insulator/superconductor model with dark matter sector.

This paper is organized as follows. In the next section, we introduce the holographic
model with dark matter sector in the background of AdS soliton away from the probe limit.
In part A of section III, we study the stability of holographic phase transitions.
And in part B and part C of section III,
we disclose properties of insulator/superconductor transitions by
examining in detail behaviors of the scalar operator and the holographic
entanglement entropy of the system.
We will summarize our main results in the last section.

\section{Equations of motion and boundary conditions}

In this work, we consider holographic phase transitions with
dark matter sector in the 5-dimensional AdS soliton spacetime. The
general Lagrange density constructed by a scalar field and two U(1)
gauge fields coupled in the gravitational background reads
\cite{LN-1}
\begin{eqnarray}\label{lagrange-1}
\mathcal{L}=R+\frac{12}{L^{2}}-\left(\frac{1}{4}F^{MN}F_{MN}+|\nabla_{M}\psi-iqA_{M}|^{2}+m^{2}\psi^{2}+\frac{1}{4}B^{MN}B_{MN}
+\frac{\alpha}{4}B^{MN}F_{MN}\right),
\end{eqnarray}
where $\psi(r)$ is the complex scalar field,
$A_{M}$ stands for the ordinary Maxwell field and another additional $U(1)$ gauge field $B_{M}$ corresponds to the dark matter field,
which is not completely decoupled with the visible matter.
$-6/L^{2}$ is the negative cosmological constant, where $L$ is the radius of AdS spactime
which will be scaled unity in the following numerical calculation.
$m$ is the mass of the scalar field,
q is the scalar charge and
$\alpha$ is the coupling parameter between the two U(1) gauge fields.

The Einstein equations for the system can be written in the form
\begin{eqnarray}\label{BHpsi}
R_{MN}-\frac{1}{2}g_{MN}R-6g_{MN}=\frac{1}{2} \tilde{T}_{MN},
\end{eqnarray}
where $\tilde{T}_{MN}$ is the energy-momentum tensor expressed as
\begin{equation}
\begin{aligned}
\tilde{ T}_{MN}=F_{M\beta}F^{\beta}_{N}+B_{M\beta}B^{\beta}_{N}+\alpha B_{M\beta}F^{\beta}_{N}+2\nabla_{M}\psi\nabla_{N}\psi
+2q^{2}A_{M}A_{N}\psi^{2}~~~~~~~~~~~~~~~~~~~~~~~~~~~\\
+g_{MN}\left(-\frac{1}{4}F_{MN}F^{MN}-\frac{1}{4}B_{MN}B^{MN}
-\frac{\alpha}{4}F_{MN}B^{MN}-\nabla_{M}\psi\nabla_{M}\psi-q^{2}A_{M}A^{M}\psi^{2}-m^{2}\psi^{2}\right).
 \end{aligned}
\end{equation}

With the variation of the considered matter fields, we obtain the
corresponding independent equations of motion in the form
\begin{eqnarray}\label{BHChi}
\nabla_{M}F^{MN}-2q^{2}A^{N}\psi^{2}+\frac{\alpha}{2}\nabla_{M}B^{MN}=0,
\end{eqnarray}
\begin{eqnarray}\label{BHChi}
\nabla_{M}\nabla^{M}\psi-q^{2}A_{M}A^{M}\psi-m^{2}\psi=0,
\end{eqnarray}
\begin{eqnarray}\label{BHChi}
\nabla_{M}B^{MN}+\frac{\alpha}{2}\nabla_{M}F^{MN}=0.
\end{eqnarray}

Putting (6) into (4), we can eliminate the dark matter field and
obtain the equation
\begin{eqnarray}\label{BHChi}
\nabla_{M}F^{MN}-\frac{2q^{2}\psi^{2}A^{N}}{\tilde{\alpha}}=0,
\end{eqnarray}
where $\tilde{\alpha}=1-\frac{\alpha^{2}}{4}$.

Substituting (7) into (6), we arrive at
\begin{eqnarray}\label{BHChi}
\nabla_{M}B^{MN}+\frac{\alpha q^{2}\psi^{2}A^{N}}{\tilde{\alpha}}=0.
\end{eqnarray}

From Eqs. (7) and (8), we find that the ordinary Maxwell field
$A_{M}$ and the $U(1)$ gauge field $B_{M}$ which corresponds to the
dark matter field share similar features for the holographic
superconductor system in AdS soliton background, so we can still use
the metric ansatz given in Ref. \cite{GTH} where the authors took
the backreaction into account (without dark matter sector), i.e.,

\begin{eqnarray}\label{AdSBH}
ds^{2}&=&-r^{2}e^{C(r)}dt^{2}+\frac{dr^{2}}{r^{2}B(r)}+r^{2}dx^{2}+r^{2}dy^{2}+r^{2}B(r)e^{D(r)}d\chi^{2},
\end{eqnarray}
\begin{eqnarray}\label{symmetryBH}
A=\phi(r)dt,~~~~~~~~B=\eta(r)dt,~~~~~~~\psi=\psi(r).
\end{eqnarray}

In order to get smooth solutions at the tip $r_{s}$ satisfying
$B(r_{s})=0$, we have to impose on the coordinate $\chi$ a period
$\Gamma$ as
\begin{eqnarray}\label{HawkingT}
\Gamma=\frac{4\pi e^{^{-D(r_{s})/2}}}{r_{s}^{2}B'(r_{s})}.
\end{eqnarray}
We also need $C(r\rightarrow\infty)=0$ and $D(r\rightarrow\infty)=0$ to recover the AdS boundary.

From above assumptions, we can obtain the equations of motion as
\begin{eqnarray}\label{BHpsi}
\psi''+\left(\frac{5}{r}+\frac{B'}{B}+\frac{C'}{2}+\frac{D'}{2}\right)\psi'+\frac{q^{2}\phi^{^{2}}
e^{-C}}{r^{4}B}\psi-\frac{m^{2}}{r^{2}B}\psi=0,
\end{eqnarray}
\begin{eqnarray}\label{BHphi}
\phi''+\left(\frac{3}{r}+\frac{B'}{B}-\frac{C'}{2}+\frac{D'}{2}\right)\phi'-\frac{2q^{2}\psi^{^{2}}}{\tilde{\alpha}r^{2}B}\phi=0,
\end{eqnarray}
\begin{eqnarray}\label{BHphi}
\eta''+\left(\frac{3}{r}+\frac{B'}{B}-\frac{C'}{2}+\frac{D'}{2}\right)\eta'+\frac{\alpha
q^{2}\psi^{2}\phi}{\tilde{\alpha}r^{2}B}=0,
\end{eqnarray}
\begin{eqnarray}\label{BHg}
C''+\frac{1}{2}C'^{2}+\left(\frac{5}{r}+\frac{A'}{2}+\frac{B'}{B}\right)C'-\frac{e^{-C}}{r^{2}}\left(\phi'^{2}+\eta'^{2}+\alpha
\phi' \eta'\right) -\frac{2q^{2}\phi^{2}\psi^{2}e^{-C}}{r^{4}B}=0,
\end{eqnarray}
\begin{eqnarray}\label{BHChi}
B'\left(\frac{3}{r}-\frac{C'}{2}\right)+B\left[\psi'^{2}-\frac{1}{2}A'C'+\frac{e^{-C}}{2r^{2}}\left(\phi'^{2}+\eta'^{2}+\alpha
\phi' \eta'\right)+\frac{12}{r^{2}}\right]+
\frac{q^{2}\phi^{2}\psi^{2}e^{-C}}{r^{4}}+\frac{m^{2}\psi^{2}}{r^{2}}-\frac{12}{r^{2}}=0,
\end{eqnarray}
\begin{eqnarray}\label{BHChi}
D'=\frac{2r^{2}C''+r^{2}C'^{2}+4rC'-2e^{-C}(\phi'^{2}+\eta'^{2}+\alpha \phi' \eta')+4r^{2}\psi'^{2}}{r(6+r C')},
\end{eqnarray}
where the prime denotes the derivative with respect to $r$. Since
the equations are nonlinear and coupled to each other, we have to
solve these equations by using the numerical shooting method which
will integrate the equations of motion from the tip of the soliton
out to the infinity. Thus, we have to specify the boundary
conditions for this system. At the tip, we can impose proper
boundary conditions as
\begin{eqnarray}\label{InfBH}
&&\psi(r)=\psi_{0}+\psi_{1}(r-r_{s})+\cdots,~~~\phi(r)=\phi_{0}+\phi_{1}(r-r_{s})+\cdots,\nonumber\\
&&\eta(r)=\eta_{0}+\eta_{1}(r-r_{s})+\cdots,~~~~~~B(r)=B_{0}(r-r_{s})+\cdots,\nonumber\\
&&C(r)=C_{0}+C_{1}(r-r_{s})+\cdots,~~~D(r)=D_{0}+D_{1}(r-r_{s})+\cdots,
\end{eqnarray}
where the dots denote higher order terms.

After putting expansions (18) into $(12)-(17)$ and considering leading terms
of these equations, we are left with six independent parameters $r_{s}$,
$\psi_{0}$, $\phi_{0}$, $\eta_{0}$, $C_{0}$ and $D_{0}$ at the tip.
Near the AdS boundary $(r\rightarrow \infty)$, the asymptotic
behaviors of the solutions are
\begin{eqnarray}\label{InfBH}
&&\psi\rightarrow\frac{\psi_{-}}{r^{\lambda_{-}}}+\frac{\psi_{+}}{r^{\lambda_{+}}}+\cdots,~~~\
\phi\rightarrow \mu-\frac{\rho}{r^{2}}+\cdots,~~~ \eta\rightarrow\xi-\frac{\varpi}{r^{2}}+\cdots,\nonumber\\
&&B\rightarrow 1+\frac{B_{4}}{r^{4}}+\cdots,~~~ C\rightarrow
\frac{C_{4}}{r^{4}}+\cdots,~~~ D\rightarrow
\frac{D_{4}}{r^{4}}+\cdots,
\end{eqnarray}
with $\lambda_{\pm}=(2\pm\sqrt{4+m^{2}})$. $\mu$ and $\rho$ can be
interpreted as the chemical potential and charge density in the dual
theory respectively. The other two operators $\xi$ and $\varpi$ are
dual to the U(1) gauge field $\eta(r)$.

From the equations of motion for the system, we obtain the scaling
symmetry
\begin{eqnarray}\label{symmetryBH}
r \rightarrow ar,~~~~~~~~(\chi,x,y,t)\rightarrow~(\chi,x,y,t)/a,~~~~~~~\phi\rightarrow
a\phi,~~~~~~~\eta\rightarrow
a\eta,
\end{eqnarray}
which can be used to set $r_{s}=1$ in the following calculation.

We will impose four constraint conditions at infinity. The
asymptotic expressions (19) imply two constraint conditions at
infinity:$C(\infty)=0$ and $D(\infty)=0$. We also have to impose one
falloff of the scalar fields vanishes in order to get a stable CFT
on the boundary. Choosing $m^{2}=-\frac{15}{4}>-4$ above the BF
bound \cite{P. Breitenlohner}, the second mode $\psi_{+}$ is always
normalizable. In this paper, we will fix $\psi_{-}=0$ and use the
operator $\psi_{+}=<O_{+}>$ to describe the phase transition in the
dual CFT. For different values of $\psi_{0}$, we can rely on the
independent parameters $\phi_{0}$, $\eta_{0}$, $C_{0}$ and $D_{0}$
as the shooting parameter to search for the solutions with the
boundary conditions $\psi_{-}=0$, $C(r\rightarrow\infty)=0$,
$D(r\rightarrow\infty)=0$ and $\frac{\xi}{\mu}$ fixed.

\section{Holographic phase transitions in AdS soliton background}

\subsection{Stability of the scalar condensation}

In this part, we investigate the stability of holographic
insulator/superconductor phase transitions through the behaviors of
the scalar operator. We choose the example $\Gamma=\pi$,
$m^{2}=-\frac{15}{4}$, $q=2$, $\alpha=2.5$ and $\frac{\xi}{\mu}=-1$
in Fig. 1. It is surprising in the left panel that the condensed
phases appear at small chemical potential $\mu <\mu_{c}=0.944$,
which is different from the normal results in \cite{TN,YQ}. We refer
this phenomenon as the retrograde condensation, which was also
observed in the holographic metal/superconductor transitions in the
background of AdS black hole \cite{LN-1,Yan Peng-0}. By choosing
different sets of parameters, we find that there are retrograde
condensations for all superconducting solutions satisfying
$\frac{\xi}{\mu}=-1$ and $\alpha>2$.

\begin{figure}[h]
\includegraphics[width=190pt]{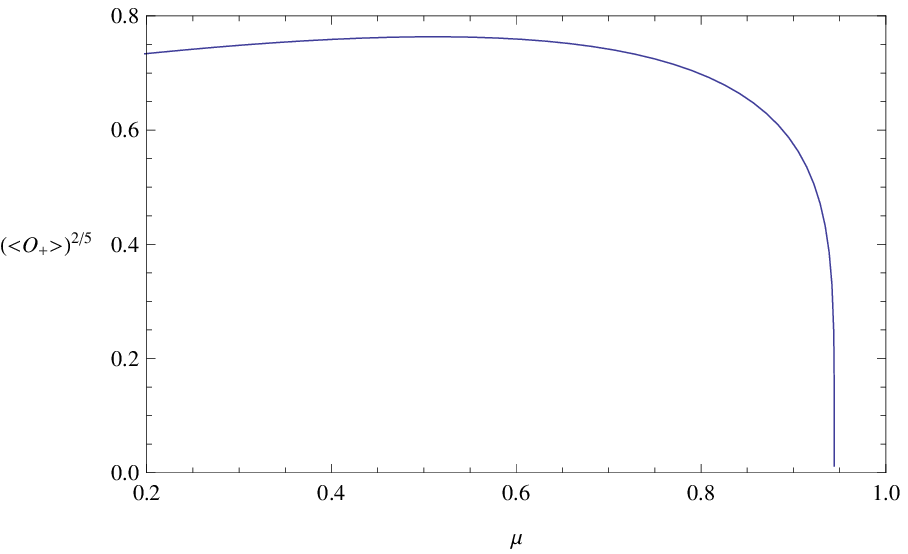}\
\includegraphics[width=175pt]{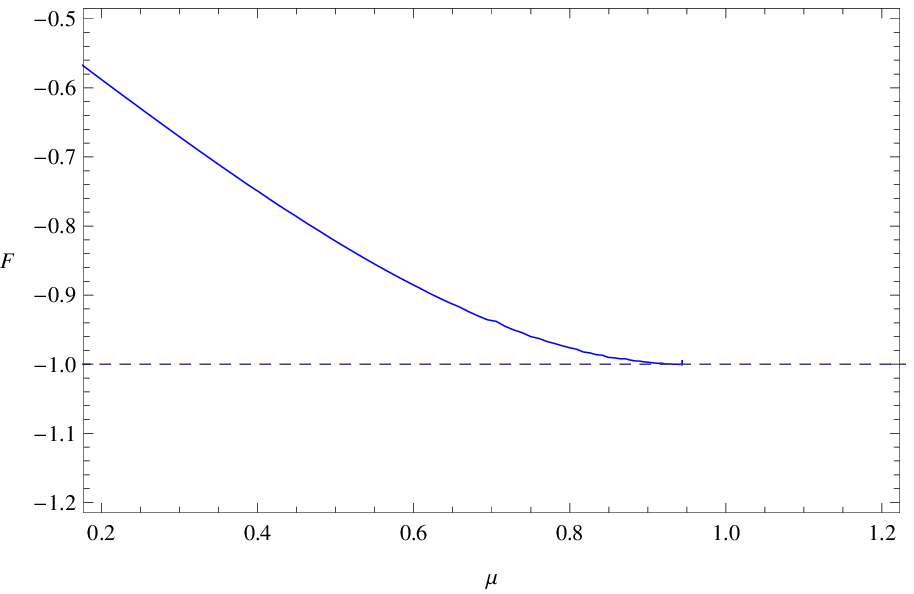}\
\caption{\label{f1} (Color online) The phase transition in cases of $\Gamma=\pi$, $m^{2}=-\frac{15}{4}$, $q=2$, $\alpha=2.5$ and $\frac{\xi}{\mu}=-1$. The left panel shows the behavior of the scalar condensation.
The right panel represents the free energy of the system,
where the solid line corresponds to the superconducting phase and the dashed line is with the normal phase.
}
\end{figure}

In order to determine whether the retrograde condensation is
thermodynamically favored, we should calculate the free energy of
the system for both normal phase and condensed phase. We show the
corresponding free energy of the system in the right panel of Fig.
1. It shows that the free energy of this hairy AdS soliton is larger than
the free energy of the soliton in normal phase. Since the physical
procedure corresponds to the phases with the lowest free energy, we
arrive at a conclusion that the retrograde condensation
superconducting solutions are thermodynamically unstable and the
unusual behaviors of the scalar operator can be used to detect the
thermodynamical instability of phase transitions.
In the retrograde condensation, the soliton superconductor phase is not thermodynamically favored.
In other words, there is no soliton/soliton superconductor transition as usual.
In this case, we expect that there maybe interesting soliton/black hole and soliton/black hole
superconductor transitions with the increase of chemical potential.
And we plan to draw the complete diagram of the soliton/black hole/black hole superconductor
system in the next work.
And on the aspects of bulk theory,
the unstable condensation also means the scalar field can't condense in the background of AdS soliton,
which corresponds to an approach from the CFT on the boundary to the AdS gravity in the bulk.

Now we study the case of $\Gamma=\pi$, $m^{2}=-\frac{15}{4}$, $q=2$, $\alpha=0.5$ and
$\frac{\xi}{\mu}=1$ in Fig. 2. In the left panel, we find a critical
chemical potential $\mu_{c}=0.944$ above which there is scalar
condensation. In order to further study the phase transition, we
plot the free energy in the right panel of Fig. 2. The free energy
of superconducting state lies below the free energy of the normal
state, which suggests that the superconducting solutions are
thermodynamically stable. More calculations show that phase transitions are
thermodynamically stable for all $\alpha>0$ and $\frac{\xi}{\mu}=1$. When $\frac{\xi}{\mu}=-1$, there are
thermodynamically stable solutions for $0\leqslant\alpha<2$.

\begin{figure}[h]
\includegraphics[width=190pt]{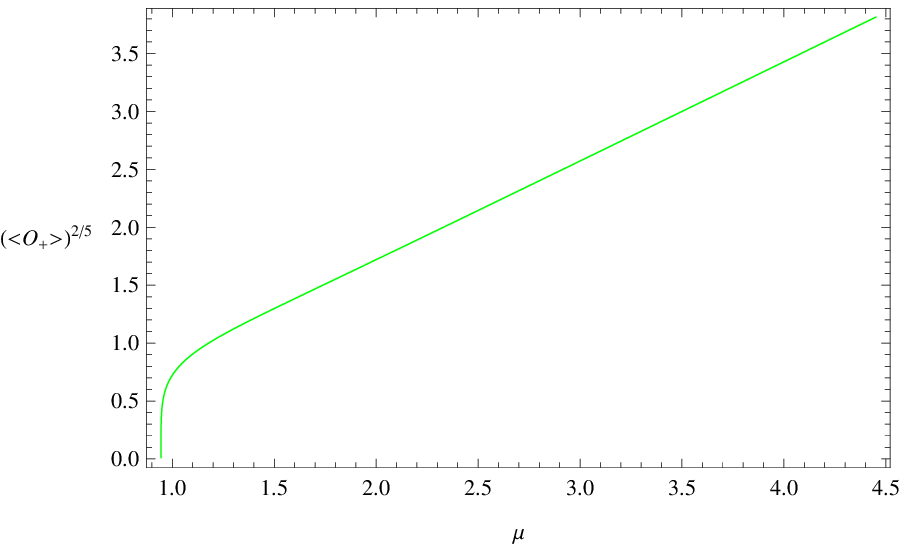}\
\includegraphics[width=175pt]{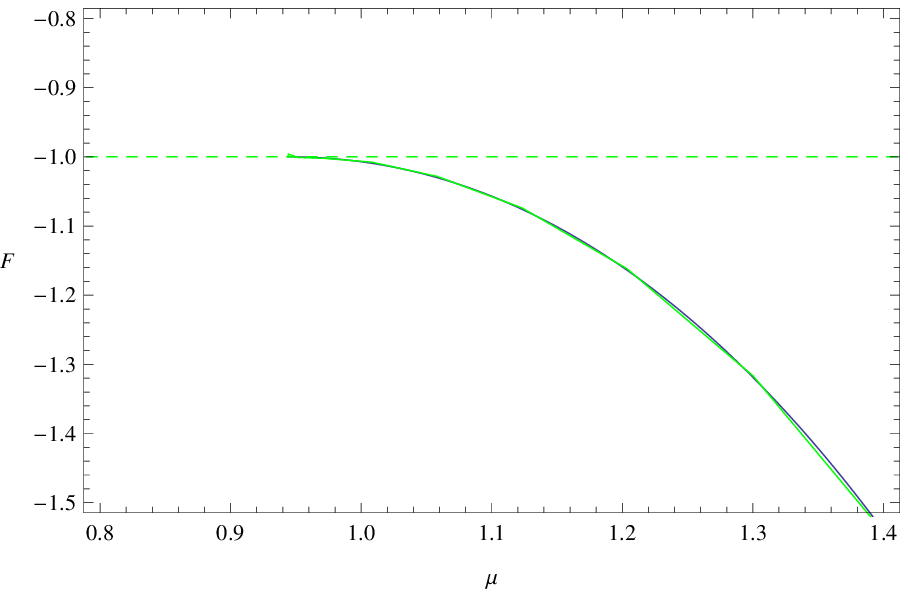}\
\caption{\label{EEntropySoliton} (Color online) The phase transitions in cases of $\Gamma=\pi$, $m^{2}=-\frac{15}{4}$, $q=2$, $\alpha=0.5$ and $\frac{\xi}{\mu}=1$.
The left panel represents the scalar operator with respect to the chemical potential.
The solid green line in the right panel shows the free energy of the superconducting phase
and the dashed line corresponds to normal phases. As a comparison, we also plot the results of equation (21)
with solid blue line in the right panel.
}
\end{figure}

With fitting methods, we obtain approximate formulas
for the free energy of normal state ($F_{SL}$) and superconducting state ($F_{SC}$) that:
\begin{eqnarray}\label{BHpsi}
F_{SL}\thickapprox -1.00,~~~~
F_{SC}\thickapprox -1.00-2.21(\mu-\mu_{c})^2-0.86(\mu-\mu_{c})^3.
\end{eqnarray}
In the right panel, the solid blue line corresponding to formula (21) almost coincides with the solid
green line obtained from the original data.
That means our fitting formula is valid.
More importantly, the front formulas suggest that: $F_{SL}|_{\mu=\mu_{c}}=F_{SC}|_{\mu=\mu_{c}}$, $\frac{\partial F_{SL}}{\partial \mu}|_{\mu=\mu_{c}}=\frac{\partial F_{SC}}{\partial \mu}|_{\mu=\mu_{c}}$, $\frac{\partial^{2} F_{SL}}{\partial \mu^{2}}|_{\mu=\mu_{c}}\neq\frac{\partial^{2} F_{SC}}{\partial \mu^{2}}|_{\mu=\mu_{c}}$.
In other words, the curve representing the physical phases with the lowest free energy is smooth at the critical point $\mu_{c}$
or the insulator/superconductor transition is of the second order.

\subsection{Properties of the scalar condensation with weak backreaction}

According to the transformation in \cite{Y. Brihaye}, larger scalar charge
q corresponds to cases of smaller backreaction parameter.
In the following discussion, we take $q=2$ as phases with weak backreaction in
this part and $q=1$ as the strong backreaction cases in the next part.
We show the scalar operator $\langle O_{+}\rangle^{1/\lambda_{+}}$ as a function
of $\mu$ with $\Gamma=\pi$, $m^{2}=-\frac{15}{4}$ and $q=2$ in Fig. 3. It is
shown that there are phase transitions at critical chemical potentials,
above which the charged scalar condensation turns on. We exhibit the condensation of the scalar operator
by choosing various $\alpha$ from top to bottom as: $\alpha=0$,
$\alpha=0.5$, $\alpha=1.0$ and $\frac{\xi}{\mu}=1$. The increase of
the parameter $\alpha$ develops shallower condensation gap.
And in both curves, the critical chemical
potential keeps as a constant $\mu_{c}=0.944$ for different values
of $\alpha$. In all, the effects of the dark
matter sector on the critical phase transition points and condensation gap are very
different from those reported in the metal/superconductor system
\cite{LN-1,LN-2,Yan Peng-0}. We conclude that the effects of the
dark matter sector on transitions depend on backgrounds.

\begin{figure}[h]
\includegraphics[width=190pt]{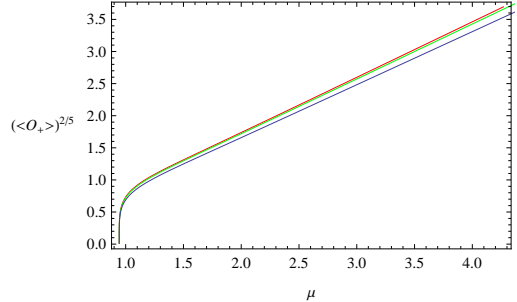}\
\caption{\label{f2} (Color online) The phase transition in the cases of $\Gamma=\pi$, $m^{2}=-\frac{15}{4}$ and $q=2$.
It shows the behavior of scalar operator with $\frac{\xi}{\mu}=1$ and various $\alpha$
from top to bottom as: $\alpha=0$ (red), $\alpha=0.5$ (green)and $\alpha=1$ (blue).
}
\end{figure}

In the following, we pay attention to the holographic entanglement entropy(HEE) of the transition system.
The authors in Refs. \cite{S-1,S-2} have provided a method to compute
the entanglement entropy of conformal field theories (CFTs) from the gravity side.
For simplicity, we consider the entanglement
entropy for a half space corresponding to a subsystem $\bar{A}$ defined by $x > 0$,
$-\frac{R}{2}<y<\frac{R}{2}$ ($R \rightarrow \infty$), $0 \leqslant \chi \leqslant \Gamma$.
Then the entanglement entropy can be expressed as \cite{RC1,RC2,W}:

\begin{eqnarray}\label{InfBH}
S_{\bar{A}}^{half}=\frac{R\Gamma}{4G_{N}}\int_{r_{0}}^{\frac{1}{\varepsilon}}re^{\frac{D(r)}{2}}dr=\frac{R\pi}{8G_{N}}(\frac{1}{\varepsilon^{2}}+S),
\end{eqnarray}
where $r=\frac{1}{\varepsilon}$ is the UV cutoff. The first term is divergent as $\varepsilon \rightarrow 0$.
In contrast, the second term does not depend on the cutoff and thus is physical important.
As a matter of fact, this finite term is the difference between the entanglement entropy in the pure
AdS soliton and the pure AdS space. $S=-1$ corresponds to the pure AdS soliton.

\begin{figure}[h]
\includegraphics[width=190pt]{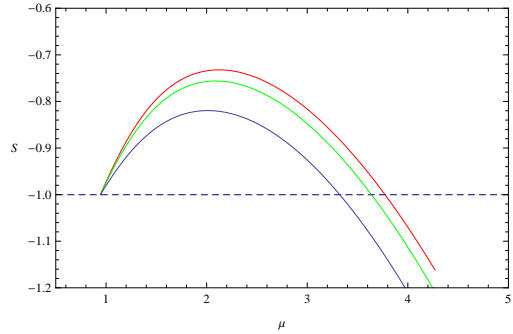}\
\caption{\label{f4} (Color online) The holographic entanglement
entropy as a function of the chemical potential $\mu$ for
$\Gamma=\pi$, $m^{2}=-\frac{15}{4}$ and $q=2$. The dashed lines
correspond to the entanglement entropy of a pure AdS soliton and the
solid lines correspond to cases of superconducting solutions. We have choose $\frac{\xi}{\mu}=1$ and various $\alpha$
from top to bottom as: $\alpha=0$ (red), $\alpha=0.5$ (green) and
$\alpha=1.0$ (blue).
}
\end{figure}

We present the holographic entanglement entropy as a function of the
chemical potential $\mu$ in Fig. 4 with $\Gamma=\pi$,
$m^{2}=-\frac{15}{4}$, $q=2$, $\frac{\xi}{\mu}=1$ and with
various $\alpha$. For all set of parameters, we find a
threshold chemical potential $\mu=0.944$, above which the hairy
soliton appears. The jump of the slop of the entanglement entropy at
$\mu=0.944$ signals that some kind of new degrees of freedom like
the Cooper pair would emerge in the new phase. It also can be easily
seen from the pictures that when the parameters are fixed, the
entanglement entropy first increases and then decreases monotonously
as we choose a larger chemical potential. It means that there is
firstly a increase and then a reduction in the number of degrees of
freedom due to the condensate generated in the phase transitions
\cite{Cai-4}. We also see that larger $\alpha$ corresponds to smaller maximum
entanglement entropy. Most
importantly, we mention that compared with the scalar operator,
the entanglement entropy is more
sensitive to the change of the coupling parameter $\alpha$.

The critical phase transition point $\mu=0.944$ obtained from the behaviors of holographic entanglement entropy is
equal to the threshold chemical potential $\mu_{c}=0.944$
obtained from the behaviors of the free energy and the scalar operator.
That means the holographic entanglement entropy can be used to search for the critical chemical potential.
With detailed analysis of the holographic entanglement
entropy, we conclude that the critical chemical potential is independent
of the dark matter sector parameter. This numerical result is nontrivial
since when considering the matter fields' backreaction on the metric, the equations
of motion depend on the coupling parameter $\alpha$ even at the phase transition
points where the scalar field is zero.
And the jump of the slope of holographic entanglement entropy
corresponds to second order phase transitions in the general holographic superconductor model with dark matter sector.
In summary, we conclude that the entanglement entropy can
be used to determine the critical phase transition point, the order
of the phase transition and the values of the coupling parameter $\alpha$.
So the entanglement entropy is indeed a good probe
of insulator/superconductor phase transitions with dark matter sector.

\subsection{Various phase transitions with strong backreaction}

Now we pay attention to the holographic superconductor with small charge $q=1$ or strong backreaction.
We exhibit the free energy as a function of the chemical potential in Fig. 5 with $\Gamma=\pi$, $m^{2}=-\frac{15}{4}$ and various $\alpha$.
It can be seen from the left panel that, for the small model parameter $\alpha=0.1$, $F$
develops a discontinuity in the first derivative of the free energy with respect to the chemical potential
at a critical value $\mu_c=1.82$, which implies the first order phase transition. In the middle panel with $\alpha=1.0$,
$F$ decreases smoothly near the critical
point $\mu_c=1.89$ indicating the second order phase transitions from normal state into superconducting state.
What's more, besides the second order phase transition at $\mu_c=1.89$,
the free energy develops a ``swallow tail'' at $\mu= 2.08$ within the superconducting phase,
a typical signal for a first order phase transition.
The right panel shows that when the coupling parameter is larger as $\alpha=1.5$,
there is only second order insulator/superconductor phase transitions at the
critical phase transition point $\mu_c=1.89$.
We conclude that the dark matter sector can affect the critical phase transition points
and also the order of phase transitions.

\begin{figure}[h]
\includegraphics[width=150pt]{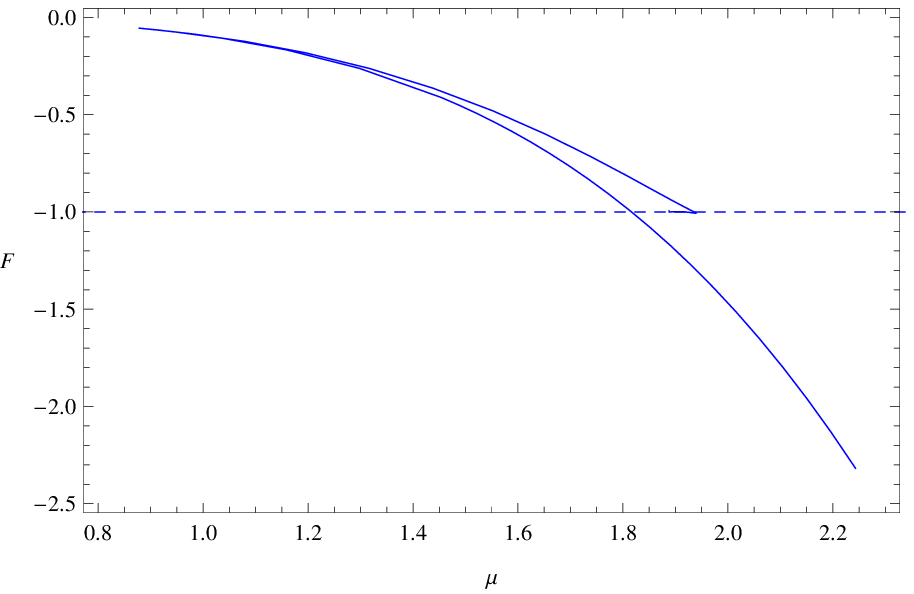}\
\includegraphics[width=150pt]{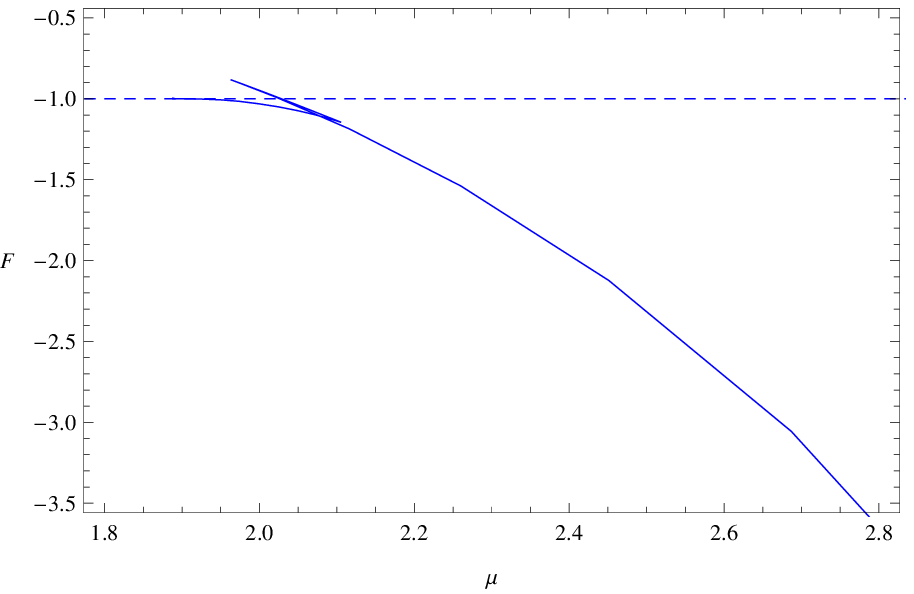}\
\includegraphics[width=150pt]{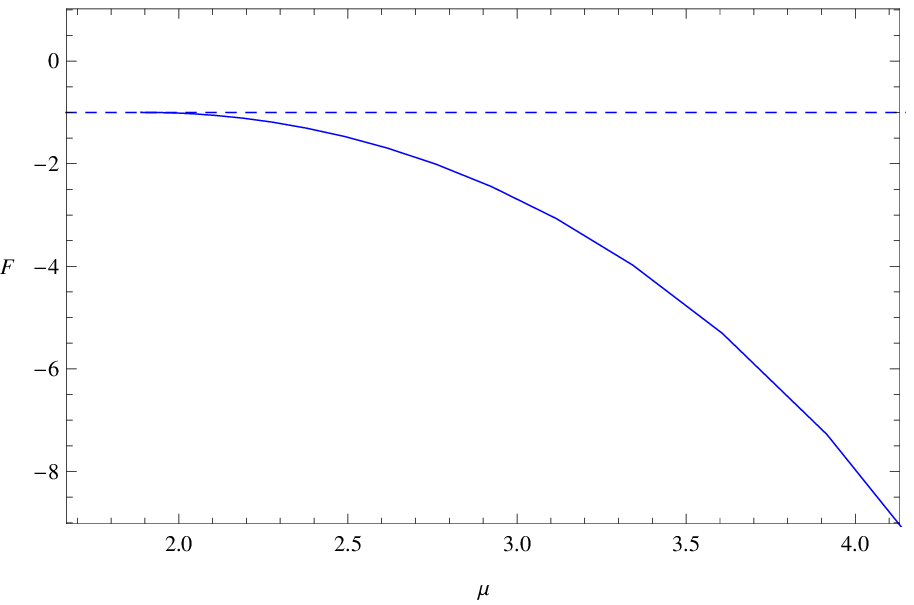}\
\caption{\label{EEntropySoliton} (Color online) The free energy with respect to the chemical potential in cases of $\Gamma=\pi$, $m^{2}=-\frac{15}{4}$, $q=1$ and $\frac{\xi}{\mu}=1$. The panels from left to right represent the cases of $\alpha=0.1$, $\alpha=1.0$ and $\alpha=1.5$. The solid line corresponds to the superconducting phase and the dashed line of $S=-1$ is with the normal phase.
}
\end{figure}

We also detect properties of phase transitions by studying the corresponding behaviors of the scalar operator in Fig. 6.
In the left panel with $\alpha=0.1$, at the critical chemical potential $\mu_{c}=1.82$,
the scalar operator $\langle O_{+}\rangle^{1/\lambda_{+}}$ has a jump
from insulator state to superconductor state indicating a first order phase transition.
When choosing $\alpha=1.0$ in the middle panel, the curves firstly increase continuously around the insulator/superconductor points $\mu_{c}=1.89$ and then have a jump at $\mu=2.08$ in the superconducting state.
It means there are second order phase transitions at $\mu_{c}=1.89$ and then first order phase transitions at $\mu=2.08$.
In cases of lager parameter $\alpha=1.5$ in the right panel, $\langle O_{+}\rangle^{1/\lambda_{+}}$ increases continuously with chemical potential, which is a classical performance of the second order phase transition.

\begin{figure}[h]
\includegraphics[width=155pt]{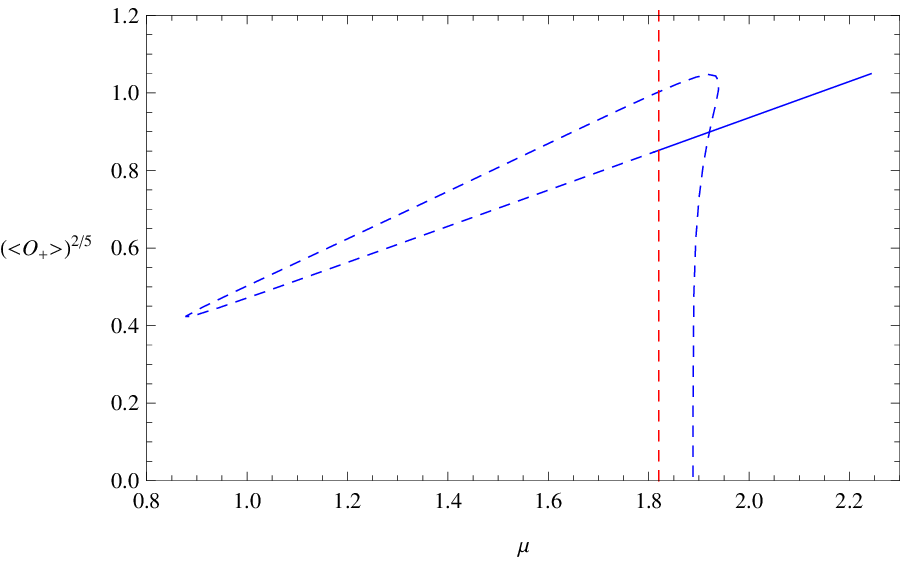}\
\includegraphics[width=155pt]{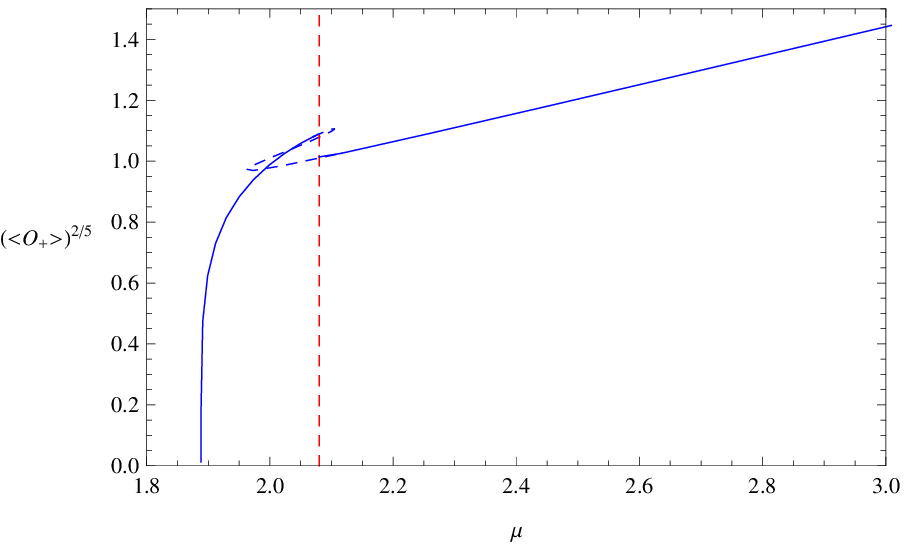}\
\includegraphics[width=155pt]{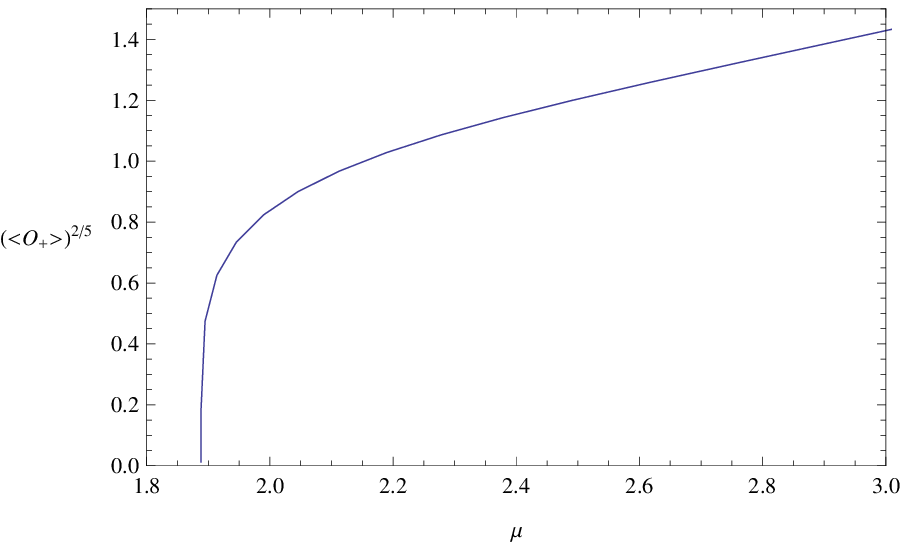}\
\caption{\label{EEntropySoliton} (Color online) The behaviors of scalar operator in cases of $\Gamma=\pi$, $m^{2}=-\frac{15}{4}$, $q=1$
and $\frac{\xi}{\mu}=1$.
The panels from left to right show the case of $\alpha=0.1$, $\alpha=1.0$ and $\alpha=1.5$. The blue solid lines correspond
to the physical superconducting states}
\end{figure}

At last, we turn to study the holographic phase transition through the entanglement entropy approach.
We plot the corresponding entanglement entropy with respect to the chemical potential $\mu$ in Fig. 7 with $\alpha=1.0$ as an example.
The blue dashed line $S=-1$ describes the entanglement entropy of the normal phase, while the blue solid line
corresponds to the entanglement entropy of the superconducting phase.
The holographic entanglement entropy is continuously at the critical chemical potential $\mu=1.89$ and there are discontinuous slops at this  transition point, which implies the phase transitions at $\mu=1.89$ are of the second order. When we go on to increase the chemical potential, the entanglement entropy has a jump around $\mu=2.08$ which corresponds to the ``swallow tail'' of the free energy $F$ in Fig. 5 and the dump of the scalar operator in Fig. 6. The front phenomenon implies a first order phase transition around $\mu=2.08$ in the superconducting phase. We state that the entanglement entropy can be used to study the critical phase transition points and the order of phase transitions in our general holographic superconductor model.
It is clear that there is numerical noise
in the picture. Since the numerical noise is small compared to the jump of the holographic entanglement
entropy at the red dashed line, it will not change our results.
\begin{figure}[h]
\includegraphics[width=190pt]{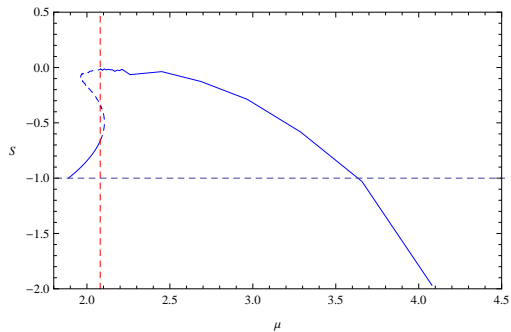}\
\caption{\label{EEntropySoliton} (Color online) The entanglement entropy in cases of $\Gamma=\pi$, $m^{2}=-\frac{15}{4}$, $q=1$, $\alpha=1.0$ and $\frac{\xi}{\mu}=1$.
The blue solid line shows the the entanglement entropy of the superconducting phase
and the blue dashed line $S=-1$ corresponds to normal phases. }
\end{figure}

\section{Conclusions}

We studied holographic insulator/superconductor
transition model with backreaction in the presence of dark matter sector.
We disclosed properties of transitions through analyzing
behaviors of the scalar operator and
the holographic entanglement entropy of the system.
For the case of weak backreaction, it was shown that the phase transition
is of the second order and the larger coupling parameter $\alpha$ makes the gap of condensation lower.
And the critical phase transition chemical potential is always a
constant for different values of $\alpha$ similar to cases without backreaction.
When including strong backreaction, the parameter $\alpha$ could affect
the critical chemical potential and the order of phase transitions,
which is very different from cases in the probe limit.
In this case, we also found the jump of the holographic topological entanglement entropy
corresponds to a first order transition in this general model with dark matter sector.
More importantly, we observed novel phases refered as retrograde condensation
due to the dark matter sector.
In the retrograde condensation, the soliton superconductor phase is not thermodynamically favored.
The front holographic properties of dark matter sector have potential application in dark matter detection
and we plan to draw the complete diagram of the soliton/black hole/black hole superconductor/soliton superconductor
system in the next work.

\begin{acknowledgments}

This work was supported by the National Natural Science Foundation
of China under Grant Nos. 11305097, 11275066 and 11505066; the
Shaanxi Province Science and Technology Department Foundation of
China under Grant Nos. 2016JQ1039 and 2016JM1028; and the Hunan
Provincial Natural Science Foundation of China under Grant No.
2016JJ1012. This work was also partly finished during the
International Conference on holographic duality for condensed matter
physics at Kavli Institute for Theoretical Physics China (KITPC),
Chinese Academy of Sciences on July 6-31, 2015.

\end{acknowledgments}

\end{document}